\documentclass[11pt]{article}
\begin{document}
\begin{flushright}
SJSU/TP-02-24\\
October 2002\end{flushright}
\vspace{1.7in}
\begin{center}\Large{\bf On ``Bell's inequality without alternative 
            settings''}\\
\vspace{1cm}
\normalsize\ J. Finkelstein\footnote[1]{
        Participating Guest, Lawrence Berkeley National Laboratory\\
        \hspace*{\parindent}\hspace*{1em}
        e-mail: JLFINKELSTEIN@lbl.gov}\\
        Department of Physics\\
        San Jos\'{e} State University\\San Jos\'{e}, CA 95192, U.S.A
\end{center}
\begin{abstract}
Cabello has recently (in quant-ph/0210081) observed that ``...an
EPR-experiment with a fixed POVM on each particle provides a violation of 
Bell's inequality without requiring local observers to choose between the
alternatives.''  In this note I discuss the implications of this 
observation for tests of locality.
\end{abstract}
\newpage
One of the difficulties for experimental tests of the Bell inequalities
\cite{Bell} is the implementation of the requirement of space-like
separation of the choice between alternative measurements on one particle
from the measurement on the other.  Cabello \cite{Cab} has recently
suggested that this requirement is not actually necessary; in fact, he
has presented a version of a Bell-violation experiment in which no such
choice is ever made.

Cabello begins by discussing the Clauser-Horne-Shimony-Holt inequality
\cite{CHSH} which he writes as
\begin{equation}
  |AB-Ab-aB-ab| \leq 2,
\end{equation}
where $A$ and $a$ are alternative measurements which might be performed
on particle 1, $B$ and $b$ are alternative measurements which might be 
performed on particle 2, and e.g. $AB$ is the expectation value of the
product of the results of $A$ and $B$.  It is well-known that, for 
suitable choice of initial two-particle state and measurement directions,
quantum theory predicts a violation of this inequality.  $A$ and $a$
cannot both be actually measured, so it is
usually assumed that a choice has to be made between $A$ and $a$
(and similarly between $B$ and $b$).  Cabello's observation is that
a single, fixed POV measurement on particle 1 can have the effect of
a measurement of $A$ or of $a$ (and a single POV measurement on particle
2 the effect of a measurement of $B$ or $b$), and so it would be possible to
perform an experiment with fixed (POV) measurements on the two particles
for which quantum theory would predict a violation of the inequality (1).
Since no choice between alternative measurements would have to be 
made, one would certainly not have to worry about making choices so
quickly that they would have space-like separation from the other 
measurement!

Suppose we consider, as an alternative to Cabello's suggestion, a
Bell-theorem experiment with projective measurements, so that a choice
between alternative settings would have to be made, but in which the choices 
are announced well in advance of the measurements.  This would not affect
the quantum predictions (given e.g.\ in eq.\ 11 of ref.\ 2), since those
predictions do not care about when the choices are made, and so quantum
theory would still predict a violation of the inequality (1).
However, a verification of this prediction would not be a proof of
nonlocality, since a local theory could agree with the quantum
prediction by allowing the measurement results on one particle to be
correlated with the setting (chosen much earlier) for the other particle.
A conflict between quantum theory and locality arises if quantum theory
predicts a violation of an inequality {\em in a case in which any local 
theory must satisfy that inequality.}

Ref.\ 2 does not offer any proof that, in the experiment proposed there,
locality would require that the inequality (1) be satisfied.
In fact, it is very easy to construct a local model which would 
exactly reproduce the quantum predictions for this experiment (as for
any experiment in which no choices are made).  The proposed POV measurements
on each particle have four possible outcomes, so there are a total of 
16 possible outcomes for each run of this experiment.  Quantum theory
predicts the probability of each of these 16 outcomes.  Just imagine that
each particle of the produced pair is given, at the moment of production,
instructions which determine which of the four possible outcomes of the
POV measurement to be performed on it is to be realized, and that the 
particles then carry the instructions to the measurement regions.
(Compare the discussion in \cite{NDM}.)
This is clearly a local model, and if the 
pairs of instructions are chosen with frequencies which
agree with the quantum probabilities, the quantum predictions will
be obeyed.  Hence a local model could agree with the quantum predictions
(and so could violate the inequality (1)) for this experiment.  Therefore
an experiment of the type suggested by Cabello should not
be considered to be a test of locality.

\vspace{1cm}
Acknowledgement: I would like to acknowledge conversations with Henry 
Stapp, as well as the hospitality of the
Lawrence Berkeley National Laboratory, where this work was done.

\end{document}